\documentclass[aps,prd,preprint,tightenlines,superscriptaddress,showpacs,byrevtex]{revtex4}
\usepackage{graphicx}
\usepackage{dcolumn}
\usepackage{bm}
\usepackage{pstricks}
\usepackage{pst-node}

\begin{document}

\preprint{\vbox{
                 \hbox{BELLE-CONF-0475}
                 \hbox{Contributed to FPCP2004}
}}

\def\bz{{B^0}}
\def\bzb{{\overline{B}{}^0}}
\def\bp{{B^+}}
\def\bm{{B^-}}
\def\kl{K_L^0}
\def\dE{{\Delta E}}
\def\mb{{M_{\rm bc}}}
\def\Dt{\Delta t}
\def\Dz{\Delta z}
\def\fol{f_{\rm ol}}
\def\fsig{f_{\rm sig}}
\newcommand{\sinbb}{{\sin2\phi_1}}
 
\newcommand{\ra}{\rightarrow}
\newcommand{\myindent}{\hspace*{2cm}}  
\newcommand{\fCP}{f_{CP}}
\def\fcp{\fCP}
\newcommand{\ftag}{f_{\rm tag}}
\newcommand{\zCP}{z_{CP}}
\newcommand{\tCP}{t_{CP}}
\newcommand{\ttag}{t_{\rm tag}}
\newcommand{\cala}{{\cal A}}
\newcommand{\calb}{{\cal B}}
\newcommand{\cals}{{\cal S}}
\newcommand{\dm}{\Delta m_d}
\newcommand{\dmd}{\dm}
\def\taubz{{\tau_\bz}}
\def\taubp{{\tau_\bp}}
\def\ks{{K_S^0}}
\newcommand{\btosqq}{b \to s\overline{q}q}
\newcommand{\btosss}{b \to s\overline{s}s}
\newcommand*{\dwl}{\ensuremath{{\Delta w_l}}}
\newcommand*{\fq}{\ensuremath{q}}
\def\kz{{K^0}}
\def\kp{{K^+}}
\def\km{{K^-}}
\def\fzero{{f_0(980)}}
\def\pip{{\pi^+}}
\def\pim{{\pi^-}}
\def\piz{{\pi^0}}
\def\kstarz{{K^{*0}}}
\def\kstarp{{K^{*+}}}
\def\kstarm{{K^{*-}}}
\def\kstarpm{{K^{*\pm}}}
\def\kl{{K_L^0}}
\def\bbar{{\overline{B}}}
\def\ufs{{\Upsilon(4S)}}
\def\nev{{N_{\rm ev}}}
\def\nsig{{N_{\rm sig}}}
\def\Nev{\nev}
\def\nsigmc{{N_{\rm sig}^{\rm MC}}}
\def\nbkg{{N_{\rm bkg}}}

\def\jpsi{{J/\psi}}
\def\dminus{{D^-}}
\def\dplus{{D^+}}
\def\dsm{{D^{*-}}}
\def\dpm{{D^{*+}}}
\def\rhop{{\rho^+}}
\def\rhom{{\rho^-}}
\def\rhoz{{\rho^0}}
\def\dzero{{D^0}}
\def\dzerob{{\overline{D}{}^0}}

\def\dzb{{\overline{D}{}^0}}
\newcommand{\dslnu}{D^{*-}\ell^+\nu}
\newcommand{\bzdslnu}{\bz \to \dslnu}
\newcommand*{\fflv}{\ensuremath{{f_\textrm{flv}}}}
\newcommand{\thetabdl}{\theta_{B,D^*\ell}}
\newcommand{\cosbdl}{\cos\thetabdl}

\def\sperp{{S_{\perp}}}
\def\lsig{{\cal L}_{\rm sig}}
\def\lbkg{{\cal L}_{\rm bkg}}
\def\rsigbkg{{\cal R}_{\rm s/b}}
\def\calf{{\cal F}}
\def\rkpi{{\cal R}_{K/\pi}}

\def\mgg{M_{\gamma\gamma}}
\def\ppizcms{p_\piz^{\rm cms}}

\def\pbstar{p_B^{\rm cms}}

\newcommand*{\eeff}{\ensuremath{\epsilon_\textrm{eff}}}
\def\egcms{{E_\gamma^{\rm cms}}}

\def\lnsig{{\cal L}_{N_{\rm sig}}}
\def\lzero{{\cal L}_0}

\def\acpraw{{A_{CP}^{\rm raw}}}

\def\kspm{{K_S^{+-}}}
\def\kszz{{K_S^{00}}}

\def\efftot{{0.30\pm 0.01}}

\def\sinbbWA{+0.726}
\def\sinbbERR{0.037}
\def\sinbbWAResult{\sinbbWA\pm\sinbbERR}

\def\SphiksResPrv{-0.96\pm0.50^{+0.09}_{-0.11}}
\def\AphiksResPrv{-0.15\pm0.29\pm0.07}
\def\SetapksResPrv{+0.43\pm0.27\pm0.05}
\def\AetapksResPrv{-0.01\pm0.16\pm0.04}
\def\SkpkmksResPrv{-0.51\pm0.26\pm0.05}
\def\AkpkmksResPrv{-0.17\pm0.16\pm0.04}


\def\NBevksksks{167} 
\def\PBksksks{0.53} \def\NBsigksksks{88\pm 13}

\def\NBevkspmkspmkspm{128} 
\def\PBkspmkspmkspm{0.56} \def\NBsigkspmkspmkspm{72\pm 10}

\def\NBevkspmkspmkszz{39} 
\def\PBkspmkspmkszz{0.40} \def\NBsigkspmkspmkszz{16\pm 8}

\def\NAevksksks{117} 
\def\PAksksks{0.54} \def\NAsigksksks{63\pm 9}

\def\NAevkspmkspmkspm{96} 
\def\PAkspmkspmkspm{0.56} \def\NAsigkspmkspmkspm{54\pm 8}

\def\NAevkspmkspmkszz{21} 
\def\PAkspmkspmkszz{0.40} \def\NAsigkspmkspmkszz{8\pm 4}

\def\Nevkspizgm{227} 
\def\Pkspizgm{0.50} \def\Nsigkspizgm{105 \pm 14}

\def\SksksksVal{+1.26} \def\SksksksStat{0.68} \def\SksksksSyst{0.18}
\def\AksksksVal{+0.54} \def\AksksksStat{0.34} \def\AksksksSyst{0.08}

\def\SkspizgmVal{-0.58} \def\SkspizgmStat{^{+0.46}_{-0.38}} \def\SkspizgmSyst{0.11}
\def\AkspizgmVal{+0.03} \def\AkspizgmStat{0.34} \def\AkspizgmSyst{0.11}

\def\SksksksResult{\SksksksVal\pm\SksksksStat\pm\SksksksSyst}
\def\SksksksResultSS
  {\SksksksVal\pm\SksksksStat\mbox{(stat)}\pm\SksksksSyst\mbox{(syst)}}

\def\AksksksResult{\AksksksVal\pm\AksksksStat\pm\AksksksSyst}
\def\AksksksResultSS
  {\AksksksVal\pm\AksksksStat\mbox{(stat)}\pm\AksksksSyst\mbox{(syst)}}

\def\SkspizgmResult{\SkspizgmVal\SkspizgmStat\pm\SkspizgmSyst}
\def\SkspizgmResultSS
  {\SkspizgmVal\SkspizgmStat\mbox{(stat)}\pm\SkspizgmSyst\mbox{(syst)}}

\def\AkspizgmResult{\AkspizgmVal\pm\AkspizgmStat\pm\AkspizgmSyst}
\def\AkspizgmResultSS
  {\AkspizgmVal\pm\AkspizgmStat\mbox{(stat)}\pm\AkspizgmSyst\mbox{(syst)}}

\def\SbsqqNewVal{+0.39} \def\SbsqqNewErr{0.11}
\def\SbsqqNewResult{\SbsqqNewVal\pm\SbsqqNewErr}

\def\Nevjpsikspm{xxxx} 
\def\Pjpsikspm{0.xx} \def\Nsigjpsikspm{xxxx\pm xx}
 \def\Nevjpsikszz{xxx} 
\def\Pjpsikszz{0.xx} \def\Nsigjpsikszz{xxx\pm xx} 
  \def\Nevjpsikl{xxxx}     
  \def\Pjpsikl{0.xx}    \def\Nsigjpsikl{xxxx\pm xx}
    \def\Nevphiks{221}       
   \def\Pphiks{0.63}     \def\Nsigphiks{139 \pm 14}
    \def\Nevphikl{207}       
   \def\Pphikl{0.17}     \def\Nsigphikl{36 \pm 15}
\def\Nevkpkmks{718}     
  \def\Pkpkmks{0.56}    \def\Nsigkpkmks{399\pm 28}
\def\NevkspizH{298}      
   \def\PkspizH{0.55}     \def\NsigkspizH{168\pm 16}
\def\NevkspizL{499}      
   \def\PkspizL{0.17}     \def\NsigkspizL{83\pm 18}
\def\Nevkstarzgm{92} 
\def\Pkstarzgm{0.65}  \def\Nsigkstarzgm{57 \pm 9}
\def\Nevetapks{842}   
  \def\Petapks{0.61}    \def\Nsigetapks{512 \pm 27}
\def\Nevomegaks{56}  
 \def\Pomegaks{0.56}    \def\Nsigomegaks{31 \pm 7}
\def\Nevfzeroetcks{178}  
 \def\Pfzeroetcks{0.58}    \def\Nsigfzeroetcks{102 \pm 12}
 \def\Pfzeroks{0.53}    \def\Nsigfzeroks{94 \pm 14}
%
\def\SjpsikzVal{+0.666}   \def\SjpsikzStat{0.046}   \def\SjpsikzSyst{x.xx}
\def\AjpsikzVal{+0.023}   \def\AjpsikzStat{0.031}   \def\AjpsikzSyst{x.xx}
\def\SphikzVal{+0.06}    \def\SphikzStat{0.33}    \def\SphikzSyst{0.09}
\def\AphikzVal{+0.08}    \def\AphikzStat{0.22}    \def\AphikzSyst{0.09}
\def\SphiksVal{-x.xx}    \def\SphiksStat{x.xx}    \def\SphiksSyst{x.xx}
\def\AphiksVal{-x.xx}    \def\AphiksStat{x.xx}    \def\AphiksSyst{x.xx}
\def\SphiklVal{-x.xx}    \def\SphiklStat{x.xx}    \def\SphiklSyst{x.xx}
\def\AphiklVal{-x.xx}    \def\AphiklStat{x.xx}    \def\AphiklSyst{x.xx}
\def\SkpkmksVal{-0.49}   \def\SkpkmksStat{0.18}   \def\SkpkmksSyst{0.04}
\def\AkpkmksVal{-0.08}   \def\AkpkmksStat{0.12}   \def\AkpkmksSyst{0.07}
\def\SkpkmksFCP{^{+0.18}_{-0.00}}
\def\SkspizVal{+0.30}    \def\SkspizStat{0.59}    \def\SkspizSyst{0.11}
\def\AkspizVal{-0.12}    \def\AkspizStat{0.20}    \def\AkspizSyst{0.07}
\def\SkstarzgmVal{-0.79} \def\SkstarzgmStat{^{+0.63}_{-0.50}} \def\SkstarzgmSyst{0.10}
\def\AkstarzgmVal{-0.00} \def\AkstarzgmStat{0.38} \def\AkstarzgmSyst{x.xx}
\def\SetapksVal{+0.65}   \def\SetapksStat{0.18}   \def\SetapksSyst{0.04}
\def\AetapksVal{-0.19}   \def\AetapksStat{0.11}   \def\AetapksSyst{0.05}
\def\SomegaksVal{+0.75}  \def\SomegaksStat{0.64}  \def\SomegaksSyst{^{+0.13}_{-0.16}}
\def\AomegaksVal{+0.26}  \def\AomegaksStat{0.48}  \def\AomegaksSyst{0.15}
\def\SfzeroksVal{+0.47}  \def\SfzeroksStat{0.41}  \def\SfzeroksSyst{0.08}
\def\AfzeroksVal{-0.39}  \def\AfzeroksStat{0.27}  \def\AfzeroksSyst{0.08}
\def\SbsqqVal{+0.43} \def\SbsqqErr{^{+0.12}_{-0.11}}

\def\SjpsikzResult{\SjpsikzVal\pm\SjpsikzStat\pm\SjpsikzSyst}
\def\SjpsikzResultSS
  {\SjpsikzVal\pm\SjpsikzStat\mbox{(stat)}\pm\SjpsikzSyst\mbox{(syst)}}
\def\AjpsikzResult{\AjpsikzVal\pm\AjpsikzStat\pm\AjpsikzSyst}
\def\AjpsikzResultSS
  {\AjpsikzVal\pm\AjpsikzStat\mbox{(stat)}\pm\AjpsikzSyst\mbox{(syst)}}
\def\SphikzResult{\SphikzVal\pm\SphikzStat\pm\SphikzSyst}
\def\SphikzResultSS
  {\SphikzVal\pm\SphikzStat\mbox{(stat)}\pm\SphikzSyst\mbox{(syst)}}
\def\AphikzResult{\AphikzVal\pm\AphikzStat\pm\AphikzSyst}
\def\AphikzResultSS
  {\AphikzVal\pm\AphikzStat\mbox{(stat)}\pm\AphikzSyst\mbox{(syst)}}
\def\SphiksResult{\SphiksVal\pm\SphiksStat\pm\SphiksSyst}
\def\SphiksResultSS
  {\SphiksVal\pm\SphiksStat\mbox{(stat)}\pm\SphiksSyst\mbox{(syst)}}
\def\AphiksResult{\AphiksVal\pm\AphiksStat\pm\AphiksSyst}
\def\AphiksResultSS
  {\AphiksVal\pm\AphiksStat\mbox{(stat)}\pm\AphiksSyst\mbox{(syst)}}
\def\SphiklResult{\SphiklVal\pm\SphiklStat\pm\SphiklSyst}
\def\SphiklResultSS
  {\SphiklVal\pm\SphiklStat\mbox{(stat)}\pm\SphiklSyst\mbox{(syst)}}
\def\AphiklResult{\AphiklVal\pm\AphiklStat\pm\AphiklSyst}
\def\AphiklResultSS
  {\AphiklVal\pm\AphiklStat\mbox{(stat)}\pm\AphiklSyst\mbox{(syst)}}
\def\SkpkmksResult{\SkpkmksVal\pm\SkpkmksStat\pm\SkpkmksSyst}
\def\SkpkmksResultSS
  {\SkpkmksVal\pm\SkpkmksStat\mbox{(stat)}\pm\SkpkmksSyst\mbox{(syst)}}
\def\AkpkmksResult{\AkpkmksVal\pm\AkpkmksStat\pm\AkpkmksSyst}
\def\AkpkmksResultSS
  {\AkpkmksVal\pm\AkpkmksStat\mbox{(stat)}\pm\AkpkmksSyst\mbox{(syst)}}
\def\SkspizResult{\SkspizVal\pm\SkspizStat\pm\SkspizSyst}
\def\SkspizResultSS
  {\SkspizVal\pm\SkspizStat\mbox{(stat)}\pm\SkspizSyst\mbox{(syst)}}
\def\AkspizResult{\AkspizVal\pm\AkspizStat\pm\AkspizSyst}
\def\AkspizResultSS
  {\AkspizVal\pm\AkspizStat\mbox{(stat)}\pm\AkspizSyst\mbox{(syst)}}
\def\SkstarzgmResult{\SkstarzgmVal\SkstarzgmStat\pm\SkstarzgmSyst}
\def\SkstarzgmResultSS
  {\SkstarzgmVal\SkstarzgmStat\mbox{(stat)}\pm\SkstarzgmSyst\mbox{(syst)}}
\def\AkstarzgmResult{\AkstarzgmVal\pm\AkstarzgmStat\pm\AkstarzgmSyst}
\def\AkstarzgmResultSS
  {\AkstarzgmVal\pm\AkstarzgmStat\mbox{(stat)}\pm\AkstarzgmSyst\mbox{(syst)}}
\def\SetapksResult{\SetapksVal\pm\SetapksStat\pm\SetapksSyst}
\def\SetapksResultSS
  {\SetapksVal\pm\SetapksStat\mbox{(stat)}\pm\SetapksSyst\mbox{(syst)}}
\def\AetapksResult{\AetapksVal\pm\AetapksStat\pm\AetapksSyst}
\def\AetapksResultSS
  {\AetapksVal\pm\AetapksStat\mbox{(stat)}\pm\AetapksSyst\mbox{(syst)}}
\def\SomegaksResult{\SomegaksVal\pm\SomegaksStat\SomegaksSyst}
\def\SomegaksResultSS
  {\SomegaksVal\pm\SomegaksStat\mbox{(stat)}\SomegaksSyst\mbox{(syst)}}
\def\AomegaksResult{\AomegaksVal\pm\AomegaksStat\pm\AomegaksSyst}
\def\AomegaksResultSS
  {\AomegaksVal\pm\AomegaksStat\mbox{(stat)}\pm\AomegaksSyst\mbox{(syst)}}
\def\SfzeroksResult{\SfzeroksVal\pm\SfzeroksStat\pm\SfzeroksSyst}
\def\SfzeroksResultSS
  {\SfzeroksVal\pm\SfzeroksStat\mbox{(stat)}\pm\SfzeroksSyst\mbox{(syst)}}
\def\AfzeroksResult{\AfzeroksVal\pm\AfzeroksStat\pm\AfzeroksSyst}
\def\AfzeroksResultSS
  {\AfzeroksVal\pm\AfzeroksStat\mbox{(stat)}\pm\AfzeroksSyst\mbox{(syst)}}
\def\SbsqqResult{\SbsqqVal\SbsqqErr}

\title{\quad\\[0.5cm] \boldmath New Measurements of Time-Dependent 
{\boldmath $CP$}-Violating\\ Asymmetries
in $\bz\to\ks\ks\ks$ and $\ks\piz\gamma$ Decays at Belle}

\date{\today}

\affiliation{Aomori University, Aomori}
\affiliation{Budker Institute of Nuclear Physics, Novosibirsk}
\affiliation{Chiba University, Chiba}
\affiliation{Chonnam National University, Kwangju}
\affiliation{Chuo University, Tokyo}
\affiliation{University of Cincinnati, Cincinnati, Ohio 45221}
\affiliation{University of Frankfurt, Frankfurt}
\affiliation{Gyeongsang National University, Chinju}
\affiliation{University of Hawaii, Honolulu, Hawaii 96822}
\affiliation{High Energy Accelerator Research Organization (KEK), Tsukuba}
\affiliation{Hiroshima Institute of Technology, Hiroshima}
\affiliation{Institute of High Energy Physics, Chinese Academy of Sciences, Beijing}
\affiliation{Institute of High Energy Physics, Vienna}
\affiliation{Institute for Theoretical and Experimental Physics, Moscow}
\affiliation{J. Stefan Institute, Ljubljana}
\affiliation{Kanagawa University, Yokohama}
\affiliation{Korea University, Seoul}
\affiliation{Kyoto University, Kyoto}
\affiliation{Kyungpook National University, Taegu}
\affiliation{Swiss Federal Institute of Technology of Lausanne, EPFL, Lausanne}
\affiliation{University of Ljubljana, Ljubljana}
\affiliation{University of Maribor, Maribor}
\affiliation{University of Melbourne, Victoria}
\affiliation{Nagoya University, Nagoya}
\affiliation{Nara Women's University, Nara}
\affiliation{National Central University, Chung-li}
\affiliation{National Kaohsiung Normal University, Kaohsiung}
\affiliation{National United University, Miao Li}
\affiliation{Department of Physics, National Taiwan University, Taipei}
\affiliation{H. Niewodniczanski Institute of Nuclear Physics, Krakow}
\affiliation{Nihon Dental College, Niigata}
\affiliation{Niigata University, Niigata}
\affiliation{Osaka City University, Osaka}
\affiliation{Osaka University, Osaka}
\affiliation{Panjab University, Chandigarh}
\affiliation{Peking University, Beijing}
\affiliation{Princeton University, Princeton, New Jersey 08545}
\affiliation{RIKEN BNL Research Center, Upton, New York 11973}
\affiliation{Saga University, Saga}
\affiliation{University of Science and Technology of China, Hefei}
\affiliation{Seoul National University, Seoul}
\affiliation{Sungkyunkwan University, Suwon}
\affiliation{University of Sydney, Sydney NSW}
\affiliation{Tata Institute of Fundamental Research, Bombay}
\affiliation{Toho University, Funabashi}
\affiliation{Tohoku Gakuin University, Tagajo}
\affiliation{Tohoku University, Sendai}
\affiliation{Department of Physics, University of Tokyo, Tokyo}
\affiliation{Tokyo Institute of Technology, Tokyo}
\affiliation{Tokyo Metropolitan University, Tokyo}
\affiliation{Tokyo University of Agriculture and Technology, Tokyo}
\affiliation{Toyama National College of Maritime Technology, Toyama}
\affiliation{University of Tsukuba, Tsukuba}
\affiliation{Utkal University, Bhubaneswer}
\affiliation{Virginia Polytechnic Institute and State University, Blacksburg, Virginia 24061}
\affiliation{Yonsei University, Seoul}
  \author{K.~Abe}\affiliation{High Energy Accelerator Research Organization (KEK), Tsukuba} 
  \author{K.~Abe}\affiliation{Tohoku Gakuin University, Tagajo} 
  \author{N.~Abe}\affiliation{Tokyo Institute of Technology, Tokyo} 
  \author{I.~Adachi}\affiliation{High Energy Accelerator Research Organization (KEK), Tsukuba} 
  \author{H.~Aihara}\affiliation{Department of Physics, University of Tokyo, Tokyo} 
  \author{M.~Akatsu}\affiliation{Nagoya University, Nagoya} 
  \author{Y.~Asano}\affiliation{University of Tsukuba, Tsukuba} 
  \author{T.~Aso}\affiliation{Toyama National College of Maritime Technology, Toyama} 
  \author{V.~Aulchenko}\affiliation{Budker Institute of Nuclear Physics, Novosibirsk} 
  \author{T.~Aushev}\affiliation{Institute for Theoretical and Experimental Physics, Moscow} 
% ...
\collaboration{The Belle Collaboration}

\begin{abstract}
  We present new measurements of $CP$-violation parameters
  in 
  $\bz\to$
  $\ks\ks\ks$
  and
  $\ks\piz\gamma$
  decays
  based on a sample of $275\times 10^6$ $B\bbar$ pairs
  collected at the $\ufs$ resonance with
  the Belle detector at the KEKB energy-asymmetric $e^+e^-$ collider.
  One neutral $B$ meson is fully reconstructed in
  one of the specified decay channels,
  and the flavor of the accompanying $B$ meson is identified from
  its decay products.
  $CP$-violation parameters
  are obtained from the asymmetries in the distributions of
  the proper-time intervals between the two $B$ decays.
  We obtain 
  \begin{eqnarray}
    \cals_{\ks\ks\ks}     &=& \SksksksResultSS,\nonumber \\
    \cala_{\ks\ks\ks}     &=& \AksksksResultSS,\nonumber \\
    \cals_{\ks\piz\gamma} &=& \SkspizgmResultSS,\nonumber \\
    \cala_{\ks\piz\gamma} &=& \AkspizgmResultSS.\nonumber
  \end{eqnarray}
  All results are preliminary.
\end{abstract}


\maketitle

\section{Introduction}
\label{sec:introduction}

In the standard model (SM), $CP$ violation arises from an
irreducible phase, the Kobayashi-Maskawa (KM) phase~\cite{Kobayashi:1973fv},
in the weak-interaction quark-mixing matrix. In
particular, the SM predicts $CP$ asymmetries in the time-dependent
rates for $\bz$ and
$\bzb$ decays to a common $CP$ eigenstate $\fCP$~\cite{bib:sanda}. 
In the decay chain $\Upsilon(4S)\to \bz\bzb \to f_{CP}f_{\rm tag}$,
where one of the $B$ mesons decays at time $t_{CP}$ to a 
final state $f_{CP}$ 
and the other decays at time $t_{\rm tag}$ to a final state  
$f_{\rm tag}$ that distinguishes between $B^0$ and $\bzb$, 
the decay rate has a time dependence
given by
\begin{equation}
\label{eq:psig}
{\cal P}(\Delta{t}) = 
\frac{e^{-|\Delta{t}|/{\taubz}}}{4{\taubz}}
\biggl\{1 + \fq\cdot 
\Bigl[ \cals\sin(\dmd\Delta{t})
   + \cala\cos(\dmd\Delta{t})
\Bigr]
\biggr\}.
\end{equation}
Here $\cals$ and $\cala$ are $CP$-violation parameters,
$\taubz$ is the $B^0$ lifetime, $\dmd$ is the mass difference 
between the two $B^0$ mass
eigenstates, $\Delta{t}$ = $t_{CP}$ $-$ $t_{\rm tag}$, and
the $b$-flavor charge $\fq$ = +1 ($-1$) when the tagging $B$ meson
is a $B^0$ ($\bzb$).
To a good approximation,
the SM predicts $\cals = -\xi_f\sin 2\phi_1$, where $\xi_f = +1 (-1)$ 
corresponds to  $CP$-even (-odd) final states, and $\cala =0$
for both $b \to c\overline{c}s$ and 
$\btosqq$ transitions.
Recent measurements of time-dependent $CP$ asymmetries in
$\bz \to J/\psi \ks$~\cite{bib:CC} and related decay 
modes, which are governed by the $b \to c\overline{c}s$ transition,
by Belle~\cite{bib:CP1_Belle,bib:BELLE-CONF-0436}
and BaBar~\cite{bib:CP1_BaBar}
already determine $\sinbb$ rather precisely;
the present world average value is 
$\sinbb = \sinbbWAResult$~\cite{bib:HFAG}.
This serves as a firm reference point for the SM.

The phenomena of $CP$ violation in the flavor-changing $b \to s$ transition
are sensitive to physics at a very high-energy scale~\cite{Akeroyd:2004mj}.
Theoretical studies indicate that
large deviations from SM expectations
are allowed for time-dependent $CP$ asymmetries in
$\bz$ meson decays~\cite{bib:lucy}.
Experimental investigations have recently been launched
at the two $B$ factories, each of which has produced more than
$10^8$ $B\bbar$ pairs.
Belle's measurements of $CP$ asymmetries using the decay modes
$\bz\to$ 
$\phi\ks$,
$\phi\kl$,
$\kp\km\ks$,
$\fzero\ks$,
$\eta'\ks$,
$\omega\ks$,
and
$\ks\piz$,
which are dominated by the $\btosqq$ transition,
yielded a value that differs from the SM expectation 
by 2.4 standard deviations when all measurements
are combined~\cite{Abe:2004xp}.
To elucidate the difference,
it is essential to examine additional modes that
may be sensitive to the same $b \to s$ penguin amplitude.

Recently it was pointed out that
in decays of the type $B^0 \to P^0Q^0X^0$,
where $P^0$, $Q^0$ and $X^0$ 
represent spin-0 neutral particles that are $CP$ eigenstates,
the final state is a $CP$ eigenstate~\cite{Gershon:2004tk}.
The $\bz\to\ks\ks\ks$ decay, which is a $\xi_f = +1$ state,
is one of the most promising
modes in this class of decays. 
The existence of this decay mode was first reported
by the Belle collaboration~\cite{Garmash:2003er}, and
recently confirmed with larger statistics
by the BaBar collaboration~\cite{Aubert:2004dr}.
Since there is no $u$ quark in the final state,
the decay is dominated by the $\btosss$ transition.
In this report, 
we describe the first measurement of
$CP$ asymmetries in the $\bz\to\ks\ks\ks$ decay
based on a 253 fb$^{-1}$ data sample that contains
$275\times 10^6$ $B\bbar$ pairs.

We also measure time-dependent $CP$ violation in
the decay $\bz\to\ks\piz\gamma$,
which is not a $CP$ eigenstate but is sensitive to
physics beyond the SM~\cite{Atwood:1997zr,Atwood:2004jj}.
Within the SM, the photon emitted from a $\bz$ ($\bzb$)
meson is dominantly right-handed (left-handed).
Therefore the polarization of the photon carries
information on the original $b$-flavor; the decay
is thus almost flavor-specific. The SM predicts
a small asymmetry $\cals \sim -2(m_s/m_b)\sinbb$, where $m_b$ ($m_s$) is
the $b$-quark ($s$-quark) mass~\cite{Atwood:2004jj}.
Any significant deviation from this expectation
would be a manifestation of physics beyond the SM.
Belle's previous measurement of time-dependent $CP$ asymmetries in
the decay $\bz\to\kstarz\gamma~(\kstarz\to\ks\piz)$~\cite{Abe:2004xp}
required that
the invariant mass of the $\ks$ and $\piz$ ($M_{\ks\piz}$)
be between 0.8 and 1.0 GeV/$c^2$ to select the $\kstarz\to\ks\piz$
decay~\cite{Atwood:1997zr}.
Recently it was pointed out that 
in decays of the type $\bz\to P^0Q^0\gamma$
(e.g. $P^0 = \ks$ and $Q^0 = \piz$),
possible new physics effects on the mixing-induced 
$CP$ violation do not depend on the resonant structure
of the $P^0Q^0$ system~\cite{Atwood:2004jj}.
In this report, we describe a new measurement of $CP$ asymmetries
extending the invariant mass range to
$0.6 < M_{\ks\piz} < 1.8$ GeV/$c^2$, which includes
the $\kstarz$ mass region, based on the new proposal.

At the KEKB energy-asymmetric 
$e^+e^-$ (3.5 on 8.0~GeV) collider~\cite{bib:KEKB},
the $\Upsilon(4S)$ is produced
with a Lorentz boost of $\beta\gamma=0.425$ nearly along
the electron beamline ($z$).
Since the $B^0$ and $\bzb$ mesons are approximately at 
rest in the $\Upsilon(4S)$ center-of-mass system (cms),
$\Delta t$ can be determined from the displacement in $z$ 
between the $f_{CP}$ and $f_{\rm tag}$ decay vertices:
$\Delta t \simeq (z_{CP} - z_{\rm tag})/(\beta\gamma c)
 \equiv \Delta z/(\beta\gamma c)$.

The Belle detector is a large-solid-angle magnetic
spectrometer that
consists of a silicon vertex detector (SVD),
a 50-layer central drift chamber (CDC), an array of
aerogel threshold Cherenkov counters (ACC),
a barrel-like arrangement of time-of-flight
scintillation counters (TOF), and an electromagnetic calorimeter
comprised of CsI(Tl) crystals (ECL) located inside
a superconducting solenoid coil that provides a 1.5~T
magnetic field.  An iron flux-return located outside of
the coil is instrumented to detect $K_L^0$ mesons and to identify
muons (KLM).  The detector
is described in detail elsewhere~\cite{Belle}.
Two inner detector configurations were used. A 2.0 cm radius beampipe
and a 3-layer silicon vertex detector (SVD-I) were used for 
a 140 fb$^{-1}$ data sample (DS-I)
containing $152\times 16^6$ $B\bbar$ pairs,
while a 1.5 cm radius beampipe, a 4-layer
silicon detector (SVD-II)~\cite{Ushiroda} 
and a small-cell inner drift chamber were used for 
an additional 113 fb$^{-1}$
data sample (DS-II) that contains $123\times 10^6$ $B\bbar$
pairs for a total of $275\times 10^6$ $B\bbar$ pairs.

\section{Event Selection, Flavor Tagging and Vertex Reconstruction}
\subsection{\boldmath $\bz\to\ks\ks\ks$}
We reconstruct the $\bz\to\ks\ks\ks$ decay
with a $\kspm\kspm\kspm$ or $\kspm\kspm\kszz$ final state,
where a $\pip\pim$ ($\piz\piz$) 
state from a $\ks$ decay is denoted as $\kspm$ ($\kszz$).
We use these notations whenever appropriate.
Although the $\kspm\kspm\kszz$ state
suffers from a larger background and a lower
vertex reconstruction efficiency, we include it because its
total branching fraction is larger than that of the $\kspm\kspm\kspm$
mode;
with $\calb(\ks\to\pip\pim) = 2/3$ and $\calb(\ks\to\piz\piz) = 1/3$,
we obtain
$\calb(3\ks\to \kspm\kspm\kspm) = (2/3)^3 = 8/27$ 
while
$\calb(3\ks\to \kspm\kspm\kszz) = 3(2/3)^2(1/3) = 12/27$. 
We do not include final states with two or three $\ks\to\piz\piz$
since their products of efficiencies and branching fractions are small.


Pairs of oppositely charged tracks that have an invariant mass
within 0.012 GeV/$c^2$
of the nominal $\ks$ mass are used to reconstruct $\ks\to\pip\pim$ decays.
The $\pip\pim$ vertex is required to be displaced from
the IP by a minimum transverse distance of 0.22~cm for
high momentum ($>1.5$ GeV/$c$) candidates and 0.08~cm for
those with momentum less than 1.5~GeV/$c$.
The direction of the pion pair momentum must also agree with
the direction defined by the IP and the vertex displacement
within 0.03 rad for high-momentum candidates, and within 0.1
rad for the remaining candidates.
When we find two good $\kspm$ candidates that satisfy the criteria
described above, we apply looser selection criteria for
the third $\kspm$ candidate:
(1) the mismatch in the $z$ direction at the $\ks$ vertex point
    for the two $\pi^{\pm}$ tracks should be less than 5 cm
    (1cm when both pions have associated SVD hits);
(2) the angle in the $r$-$\phi$ plane between the $\ks$ momentum vector and
   the direction defined by the $\ks$ and the IP should be less than 0.2
   rad for high-momentum candidates, and less than 0.4 rad for the
   remaining candidates.
   
Photons are identified as isolated ECL clusters
that are not matched to any charged track.
To select $\ks\to\piz\piz$ decays,
we reconstruct $\piz$ candidates from pairs of photons
with $E_\gamma > 0.05$ GeV, where
$E_\gamma$ is the photon energy measured with the ECL.
The reconstructed $\piz$ candidate is required to satisfy 
$0.08~{\rm GeV}/c^2 < \mgg < 0.15~{\rm GeV}/c^2$
and
$\ppizcms > 0.1~{\rm GeV}/c$, where
$\mgg$ and $\ppizcms$ are the invariant mass and
the momentum in the cms, respectively. 
The large mass range is used to achieve a high reconstruction efficiency.
Candidate $\ks\to\piz\piz$ decays are required to have
invariant masses between 0.47 GeV/$c^2$ and 0.52 GeV/$c^2$,
where we perform a fit with constraints on
the $\ks$ vertex and the $\piz$ masses
to improve the $\piz\piz$ invariant mass resolution.
We also require that the distance between the IP and the
reconstructed $\ks$ decay vertex be between $-10$~cm and
100 cm,
where the positive direction is defined by the $\ks$ momentum.
The $\kszz$ candidate is combined with two good $\kspm$ candidates
to reconstruct $\bz\to\kspm\kspm\kszz$ decay, where we only use
aforementioned good $\kspm$ candidates.

For reconstructed $B\to\ks\ks\ks$ candidates,
we identify $B$ meson decays using the
energy difference $\dE\equiv E_B^{\rm cms}-E_{\rm beam}^{\rm cms}$ and
the beam-energy constrained mass $\mb\equiv\sqrt{(E_{\rm beam}^{\rm cms})^2-
(p_B^{\rm cms})^2}$, where $E_{\rm beam}^{\rm cms}$ is
the beam energy in the cms, and
$E_B^{\rm cms}$ and $p_B^{\rm cms}$ are the cms energy and momentum of the 
reconstructed $B$ candidate, respectively.
The $B$ meson signal region is defined as 
$|\dE|<0.10$ GeV for $\bz \to \kspm\kspm\kspm$,
$-0.15~{\rm GeV} < \dE < 0.10$ GeV for $\bz \to \kspm\kspm\kszz$,
and $5.27~{\rm GeV}/c^2 <\mb<5.29~{\rm GeV}/c^2$ for both decays.

The dominant background to the $\bz\to\ks\ks\ks$ decay comes from
$e^+e^- \rightarrow 
u\overline{u},~d\overline{d},~s\overline{s}$, or $c\overline{c}$
continuum events. Since these tend to be jet-like, while
the signal events tend to be spherical,
we use a set of variables that characterize the event topology
to distinguish between the two.
We combine modified Fox-Wolfram moments~\cite{Abe:2001nq}
into a Fisher discriminant $\calf$.
We also use the angle of the reconstructed $\bz$
candidate with respect to the beam direction in the cms
($\theta_B$).
We combine
$\calf$ and $\cos\theta_B$ into a signal [background]
likelihood variable, which is defined as 
${\cal L}_{\rm sig[bkg]} \equiv
{\cal L}_{\rm sig[bkg]}(\calf)\times
{\cal L}_{\rm sig[bkg]}(\cos\theta_B)$.
We impose requirements on the likelihood ratio 
$\rsigbkg \equiv \lsig/(\lsig+\lbkg)$ to
maximize the figure-of-merit (FoM) defined as
$\nsigmc/\sqrt{\nsigmc+\nbkg}$, where $\nsigmc$ ($\nbkg$) 
represents the expected
number of signal (background) events in the signal region.
We estimate $\nsigmc$ using Monte Carlo (MC) events, while
$\nbkg$ is determined from events outside the signal region.
The requirement for $\rsigbkg$
depends both on the decay mode
and on the flavor-tagging quality, $r$, which is
described in Sec.~\ref{sec:flavor tagging}.
The threshold values range from 0.2 (used for $r>0.875$)
to 0.5 (used for $r<0.25$) for the case
with 3 good $\kspm$ candidates,
and from 0.3 to 0.9 for other cases.

We reject $\ks\ks\ks$ candidates that are consistent with
$\bz\to\chi_{c0}\ks\to(\ks\ks)\ks$ or $\bz\to D^0\ks\to(\ks\ks)\ks$.
We keep candidate $\bz\to\fzero\ks\to(\ks\ks)\ks$ decays as
they are also dominated by the $b\to s$ transition.

We use events outside the signal region 
as well as a large MC sample to study the background components.
The dominant background is from continuum.
The contributions from $B\overline{B}$ events are small.
The contamination of $\bz\to\chi_{c0}\ks$ events in the 
$\bz\to\ks\ks\ks$ sample is small ($2.6$\%).
The influence of this background
is treated as a source of systematic uncertainty.
Backgrounds from the decay $\bz\to D^0\ks$
are found to be negligible.

\begin{figure}
\resizebox{0.49\textwidth}{!}{\includegraphics{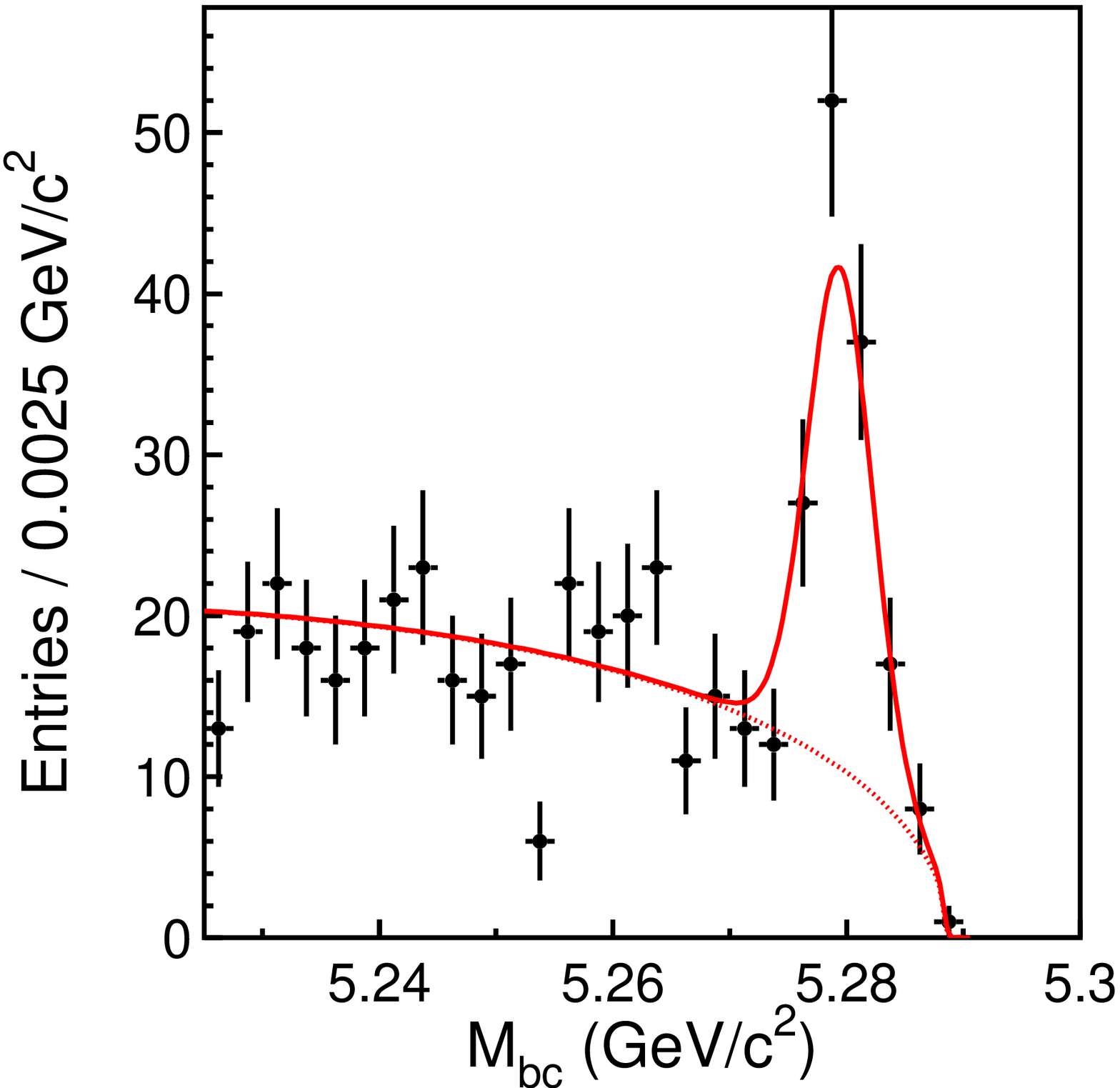}}
\resizebox{0.49\textwidth}{!}{\includegraphics{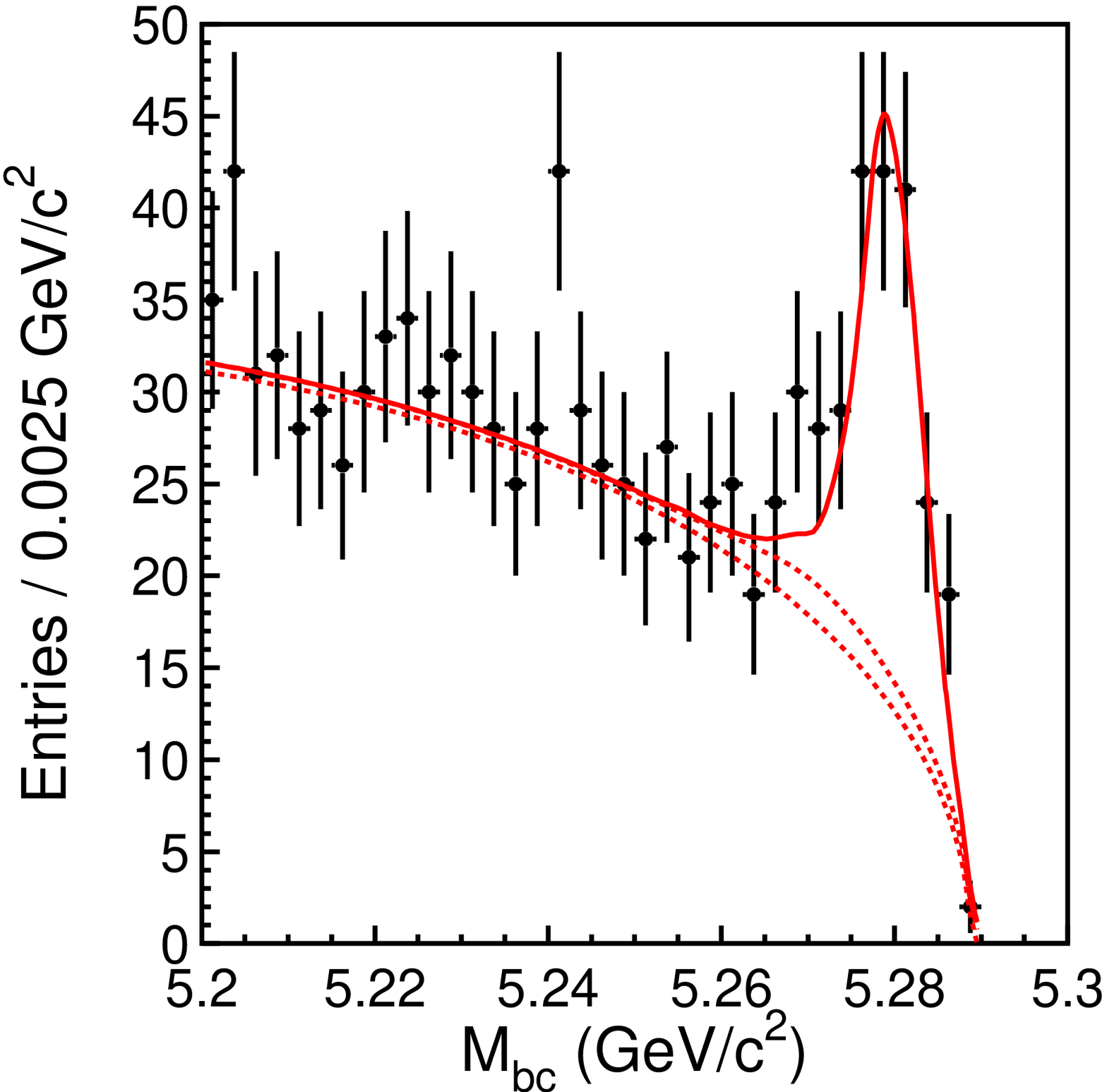}}
\resizebox{0.49\textwidth}{!}{\includegraphics{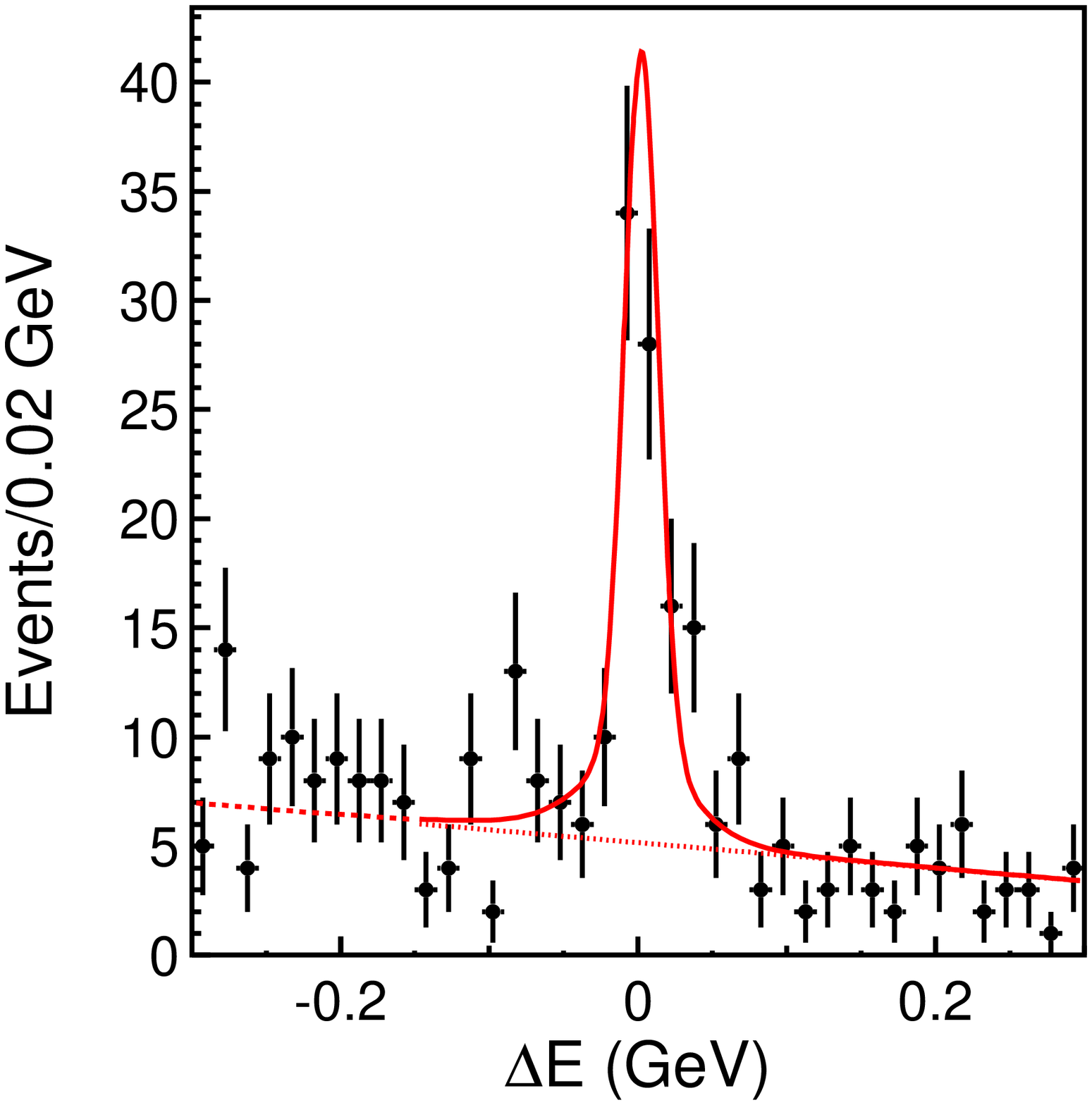}}
\resizebox{0.49\textwidth}{!}{\includegraphics{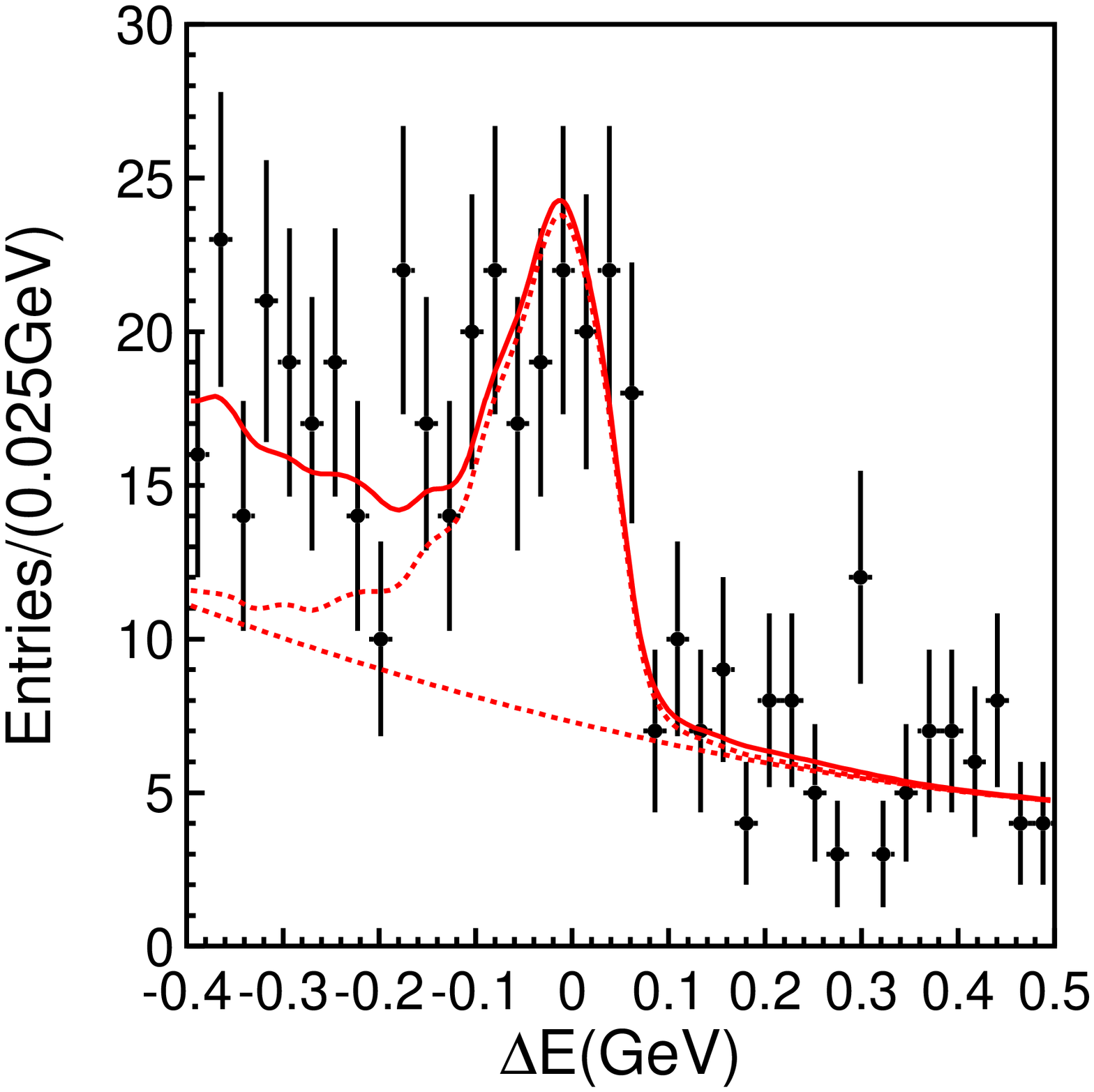}}
\caption{The $\mb$ distributions within the $\dE$ signal region for
(a) $\bz\to\ks\ks\ks$ and (b) $\bz\to\ks\piz\gamma$, 
and the $\dE$ distributions within the $\mb$ signal region for
(c) $\bz\to\ks\ks\ks$ and (d) $\bz\to\ks\piz\gamma$.
Solid curves show the fit to signal plus background distributions,
and dashed curves show the background contributions.}
\label{fig:mb}
\rput[l](-6.5, 17.5)  {(a)~$\ks\ks\ks$}
\rput[l]( 1.8, 12.8)  {(b)~$\ks\piz\gamma$}
\rput[l](-6.5,  9.5)  {(c)~$\ks\ks\ks$}
\rput[l]( 1.8,  4.8)  {(d)~$\ks\piz\gamma$}
\end{figure}

Figure~\ref{fig:mb}(a) [(c)] shows
the $\mb$ [$\dE$] distribution for the reconstructed $\bz\to\ks\ks\ks$ 
candidates within the $\dE$ [$\mb$] signal regions
after flavor tagging and vertex reconstruction.
The signal yield is determined
from an unbinned two-dimensional maximum-likelihood fit
to the $\dE$-$\mb$ distribution. The fit region
for the $\bz\to\kspm\kspm\kspm$ decay is defined as
$ 5.20~{\rm GeV/}c^2 < \mb < 5.30~{\rm GeV/}c^2$
and $-0.30~{\rm GeV} < \dE < 0.50~{\rm GeV}$,
excluding the region
$ 5.26~{\rm GeV/}c^2 < \mb < 5.30~{\rm GeV/}c^2$
and $-0.30~{\rm GeV} < \dE < -0.12~{\rm GeV}$ to
reduce the effect of background from $B$ decays.
The fit region for the $\bz\to\kspm\kspm\kszz$ decay is
defined as
$ 5.22~{\rm GeV/}c^2 < \mb < 5.30~{\rm GeV/}c^2$
and $-0.40~{\rm GeV} < \dE < 0.40~{\rm GeV}$,
excluding the region
$ 5.25~{\rm GeV/}c^2 < \mb < 5.30~{\rm GeV/}c^2$
and $-0.4~{\rm GeV} < \dE < -0.17~{\rm GeV}$.
The $\kspm\kspm\kspm$ signal distribution 
is modeled with a Gaussian function (a sum of two Gaussian functions)
for $\mb$ ($\dE$).
For the $\bz\to\kspm\kspm\kszz$ decay,
the signal is modeled with a two-dimensional
smoothed histogram obtained from MC events.
For the continuum background,
we use the ARGUS parameterization~\cite{bib:ARGUS} 
for $\mb$
and a linear function for $\dE$.
The fits after flavor tagging
yield 
$\NBsigkspmkspmkspm$ $\bz\to\kspm\kspm\kspm$ events and
$\NBsigkspmkspmkszz$ $\bz\to\kspm\kspm\kszz$ events
for a total of $\NBsigksksks$ $\bz\to\ks\ks\ks$ events 
in the signal region,
where the errors are statistical only.

\subsection{\boldmath $\bz\to\ks\piz\gamma$}
Candidate $\ks \to \pip\pim$ decays
are selected with the same criteria as those used to
select good $\kspm$ candidates for the $\bz\to\ks\ks\ks$ decay,
except that we impose
a more stringent invariant mass requirement;
only pairs of oppositely charged pions that have an invariant mass
within 0.006 GeV/$c^2$
of the nominal $\ks$ mass are used.
%
Candidate $\piz$ mesons are required to satisfy 
$0.118~{\rm GeV}/c^2 < \mgg < 0.15~{\rm GeV}/c^2$
and
$\ppizcms > 0.3~{\rm GeV}/c$.
The $\ks\piz$ invariant mass, $M_{\ks\piz}$, is required to be
between 0.6 and 1.8 GeV/$c^2$.

For prompt photons from the $\bz\to\ks\piz\gamma$ decay,
we select the photon with the largest
$\egcms$ among photon candidates in an event and require
$1.4~{\rm GeV} < \egcms < 3.4~{\rm GeV}$,
where $\egcms$ is the photon energy in the cms.
For the selected photon,
we also require $E_9/E_{25} > 0.95$, 
where $E_9/E_{25}$ is 
the ratio of energies summed in
$3\times 3$ and $5\times 5$ arrays of CsI(Tl) crystals surrounding
the crystal at the center of the shower.
Photons for candidate $\piz\to\gamma\gamma$ or 
$\eta\to\gamma\gamma$
decays are not used; we reject photon pairs that satisfy
$\mathcal{L}_{\piz}\ge 0.18$ or $\mathcal{L}_{\eta}\ge 0.18$,
where $\mathcal{L}_{\piz(\eta)}$ is a 
$\piz$ ($\eta$) likelihood described in detail
elsewhere~\cite{Koppenburg:2004fz}.
The polar angle of the
photon direction in the laboratory frame is required to be between
$33^\circ$ and $128^\circ$ for DS-I, 
while no requirement is imposed for DS-II 
as the material within the acceptance of the ECL is much reduced
for this dataset.

Candidate $\bp\to\ks\pip\gamma$ decays are also selected using a similar
procedure to reconstruct the decay
$\bz\to\ks\piz\gamma$.
Candidate $\bp\to\ks\pip\gamma$ and
$\bz\to\ks\piz\gamma$ decays are selected simultaneously;
we allow only one candidate for each event.
The best candidate selection is based on the event likelihood ratio
$\rsigbkg$ that is obtained by
combining $\calf$, which uses
the extended modified Fox-Wolfram moments~\cite{Abe:2003yy} as
discriminating variables, 
with $\cos\theta_H$ defined as the
angle between the $\bz$ meson momentum and the daughter
$\ks$ momentum in the rest frame of the $\ks\piz$ system.
We select the candidate with the largest $\rsigbkg$.

The signal region for the $\bz\to\ks\piz\gamma$ decay
is defined as 
$-0.2$ GeV $< \dE <0.1$ GeV,
$5.27~{\rm GeV}/c^2 <\mb<5.29~{\rm GeV}/c^2$.
We require $\rsigbkg > 0.5$ to reduce the continuum background.

The selection criteria described above are the same as those used
for the previous time-dependent $CP$ asymmetry
measurement in 
the $\bz\to\kstarz\gamma~(\kstarz\to\ks\piz)$ decay~\cite{Abe:2004xp},
except for the wider $M_{\ks\piz}$ range.

We use events outside the signal region 
as well as a large MC sample to study the background components.
The dominant background is from continuum.
Background contributions from $B$ decays 
are significantly smaller than those from continuum,
and are dominated by
cross-feed from 
other radiative $B$ decays including
$\bp\to\ks\pip\gamma$, and charmless $B$ decays.
Background from other $B\bbar$ decays is found to be negligible.

Figure~\ref{fig:mb}(b) [(d)] shows
the $\mb$ [$\dE$]
distribution for the reconstructed $\bz\to\ks\piz\gamma$ candidates
within the $\dE$ [$\mb$] signal region
after flavor tagging and vertex reconstruction.
The signal yield is determined
from an unbinned 
two-dimensional maximum-likelihood fit
to the $\dE$-$\mb$ distribution in the fit region defined as
$ 5.20~{\rm GeV/}c^2 < \mb < 5.29~{\rm GeV/}c^2$
and
$-0.4~{\rm GeV} < \dE < 0.5~{\rm GeV}$.
The $\bz\to\ks\piz\gamma$ signal distribution is represented by
a smoothed histogram obtained from MC simulation that
accounts for the correlation between $\mb$ and $\dE$.
The background from $B$ decays is also modeled with
a smoothed histogram obtained from MC events;
its normalization is a free parameter in
the fit. For the continuum background,
we use the ARGUS parameterization for $\mb$
and a second-order Chebyshev function for $\dE$.
The fit yields $\Nsigkspizgm$ $\bz\to\ks\piz\gamma$
events, where the error is statistical only.
For reference, we also measure the signal
before vertex reconstruction and flavor tagging, and
obtain $221\pm 21$ events.

\subsection{Flavor Tagging}
\label{sec:flavor tagging}
The $b$-flavor of the accompanying $B$ meson is identified
from inclusive properties of particles
that are not associated with the reconstructed $\bz \to \fCP$ decay~\cite{footnote:fcp}.
We use the same procedure that is used for the
$\sinbb$ measurement~\cite{bib:BELLE-CONF-0436}.
The algorithm for flavor tagging is described in detail
elsewhere~\cite{bib:fbtg_nim}.
We use two parameters, $\fq$ and $r$, to represent the tagging information.
The first, $\fq$, is already defined in Eq.~(\ref{eq:psig}).
The parameter $r$ is an event-by-event,
MC-determined flavor-tagging dilution factor
that ranges from $r=0$ for no flavor
discrimination to $r=1$ for unambiguous flavor assignment.
It is used only to sort data into six $r$ intervals.
The wrong tag fractions for the six $r$ intervals, 
$w_l~(l=1,6)$, and differences 
between $\bz$ and $\bzb$ decays, $\dwl$,
are determined from the data;
we use the same values
that were used for the $\sin 2\phi_1$ measurement~\cite{bib:BELLE-CONF-0436}
for DS-I.
Wrong tag fractions for DS-II are separately obtained 
with the same procedure; we find that the values for DS-II,
which are listed in Table~\ref{tab:wtag},
are slightly smaller than those for DS-I.
The total effective tagging efficiency for DS-II
is determined to be
$\eeff \equiv \sum_{l=1}^6 \epsilon_l(1-2w_l)^2 = \efftot$,
where $\epsilon_l$ is the event fraction for each $r$ interval
determined from the $\jpsi\ks$ data and listed in
Table~\ref{tab:wtag}.
The error includes both statistical and systematic uncertainties.

\subsection{Vertex Reconstruction}
The vertex position
for $\bz\to\ks\ks\ks$ and $\bz\to\ks\piz\gamma$ decays is
reconstructed using charged pions from $\ks$ decays.
A constraint on the IP is also used with the selected tracks;
the IP profile is convolved with finite $B$ flight length in the plane
perpendicular to the $z$ axis.
Both charged pions from the $\ks$ decay are required to
have enough SVD hits to reconstruct a $\ks$ trajectory
with a sufficient resolution.
The reconstruction efficiency depends both on
the $\ks$ momentum and on the SVD geometry.
Efficiencies with SVD-II are higher than
those with SVD-I
because of the larger outer radius and the additional layer.

The $\ftag$ vertex determination with SVD-I
remains unchanged from the
previous publication~\cite{Abe:2004xp},
and is described in detail elsewhere~\cite{bib:resol_nim};
to minimize the effect of long-lived particles, 
secondary vertices from charmed hadrons and a small fraction of
poorly reconstructed tracks, we adopt an iterative procedure
in which the track that gives the largest contribution to the
vertex $\chi^2$ is removed at each step 
until a good $\chi^2$ is obtained.

For SVD-II~\cite{Abe:2004xp}, we find that
the same vertex reconstruction algorithm results in
a larger outlier fraction when only
one track remains after the iteration procedure.
Therefore, in this case, we repeat the
iteration procedure with
a more stringent requirement on the SVD-II hit pattern.
The resulting outlier fraction is comparable to
that for SVD-I, while 
the inefficiency caused by this change is small (2.5\%).

\subsection{Summary of Signal Yields}
The signal yields for $\bz\to \fcp$ decays, $\nsig$,
after flavor tagging
are summarized in Table~\ref{tab:num}.
For the $\bz\to\ks\ks\ks$ decay, results
both before and after vertex reconstruction are listed.
The result for the $\bz\to\ks\piz\gamma$ is obtained
after vertex reconstruction.
The signal purities are also listed in the table.

\section{Results of {\boldmath $CP$} Asymmetry Measurements}
We determine $\cals$ and $\cala$ for each mode by performing an unbinned
maximum-likelihood fit to the observed $\Dt$ distribution.
The probability density function (PDF) expected for the signal
distribution, ${\cal P}_{\rm sig}(\Dt;\cals,\cala,\fq,w_l,\dwl)$, 
is given by Eq.~(\ref{eq:psig}) incorporating
the effect of incorrect flavor assignment. The distribution is
convolved with the
proper-time interval resolution function 
$R_{\rm sig}(\Dt)$,
which takes into account the finite vertex resolution. 

The $\Dt$ resolution function for the $\bz\to\ks\piz\gamma$
decay is the same as that used in the previous analysis on the decay
$\bz\to\kstarz\gamma~(\kstarz\to\ks\piz)$~\cite{Abe:2004xp}.
It is based on the resolution function obtained from
flavor-specific $B$ decays governed by
semileptonic or hadronic $b\to c$ transitions,
with additional parameters that rescale vertex
errors. The rescaling parameters depend on
the detector configuration (SVD-I or SVD-II), 
SVD hit patterns of charged pions from the $\ks$ decay,
and $\ks$ decay vertex position in the plane 
perpendicular to the beam axis.
These parameters are determined from a fit to the 
$\Dt$ distribution of $\bz\to\jpsi\ks$ data.
Here the $\ks$ and the IP constraint are used for the
vertex reconstruction, the $\bz$ lifetime is
fixed at the world average value, and $b$-flavor tagging
information is not used so that the 
expected PDF is
an exponential function convolved with
the resolution function.

We check the resulting resolution function
by also reconstructing the vertex with
leptons from $\jpsi$ decays and the IP constraint.
We find that the distribution of the
distance between the vertex positions obtained with
the two methods is well represented by
the obtained resolution function convolved with
the well-known resolution for the $\jpsi$ vertex.
Finally, we also perform a fit to the $\bz\to\jpsi\ks$ sample
with $b$-flavor information and obtain
$\cals_{\jpsi\ks} = +0.68\pm 0.10$(stat) and
$\cala_{\jpsi\ks} = +0.02\pm 0.04$(stat), which
are in good agreement with the world average values.
Thus, we conclude that 
the vertex resolution for the $\bz\to\ks\piz\gamma$
decay is well understood.

The resolution function for the $\bz\to\ks\ks\ks$
decay is based on the same resolution function
parameterization.
The rescaling factor depends on the detector configuration
(SVD-I or SVD-II),
SVD hit patterns of charged pions from the $\ks$ decay,
$\ks$ decay vertex position in the plane 
perpendicular to the beam axis, and
the number of $\kspm$ decays used for
the vertex reconstruction.

We determine the following likelihood value for each
event:
\begin{eqnarray}
P_i
&=& (1-\fol)\int \biggl[
\fsig{\cal P}_{\rm sig}(\Dt')R_{\rm sig}(\Dt_i-\Dt') \nonumber \\
&+&(1-\fsig){\cal P}_{\rm bkg}(\Dt')R_{\rm bkg}(\Dt_i-\Dt')\biggr]
d(\Dt')  \nonumber \\
&+&\fol P_{\rm ol}(\Dt_i) 
\label{eq:likelihood}
\end{eqnarray}
where $P_{\rm ol}(\Dt)$ is a broad Gaussian function that represents
an outlier component with a small fraction $\fol$.
The signal probability $\fsig$ depends on the $r$ region and
is calculated on an event-by-event basis
as a function of $\dE$ and $\mb$ for each mode.
A PDF for background events, ${\cal P}_{\rm bkg}(\Dt)$,
is modeled as a sum of exponential and prompt components, and
is convolved with a sum of two Gaussians $R_{\rm bkg}$.
All parameters in ${\cal P}_{\rm bkg} (\Dt)$
and $R_{\rm bkg}$ are determined by the fit to the $\Dt$ distribution of a 
background-enhanced control sample; i.e. events 
outside of the
$\dE$-$\mb$ signal region.
We fix $\tau_\bz$ and $\dmd$ at
their world-average values~\cite{bib:PDG2004}.
In order to reduce the statistical error on $\cala$,
we include events without vertex information
in the analysis of $\bz\to\ks\ks\ks$.
The likelihood value in this case is obtained by integrating 
Eq.~(\ref{eq:likelihood}) over $\Dt_i$.

The only free parameters in the final fit
are $\cals$ and $\cala$, which are determined by maximizing the
likelihood function
$L = \prod_iP_i(\Dt_i;\cals,\cala)$
where the product is over all events.

\begin{figure}
\resizebox{!}{0.49\textwidth}{\includegraphics{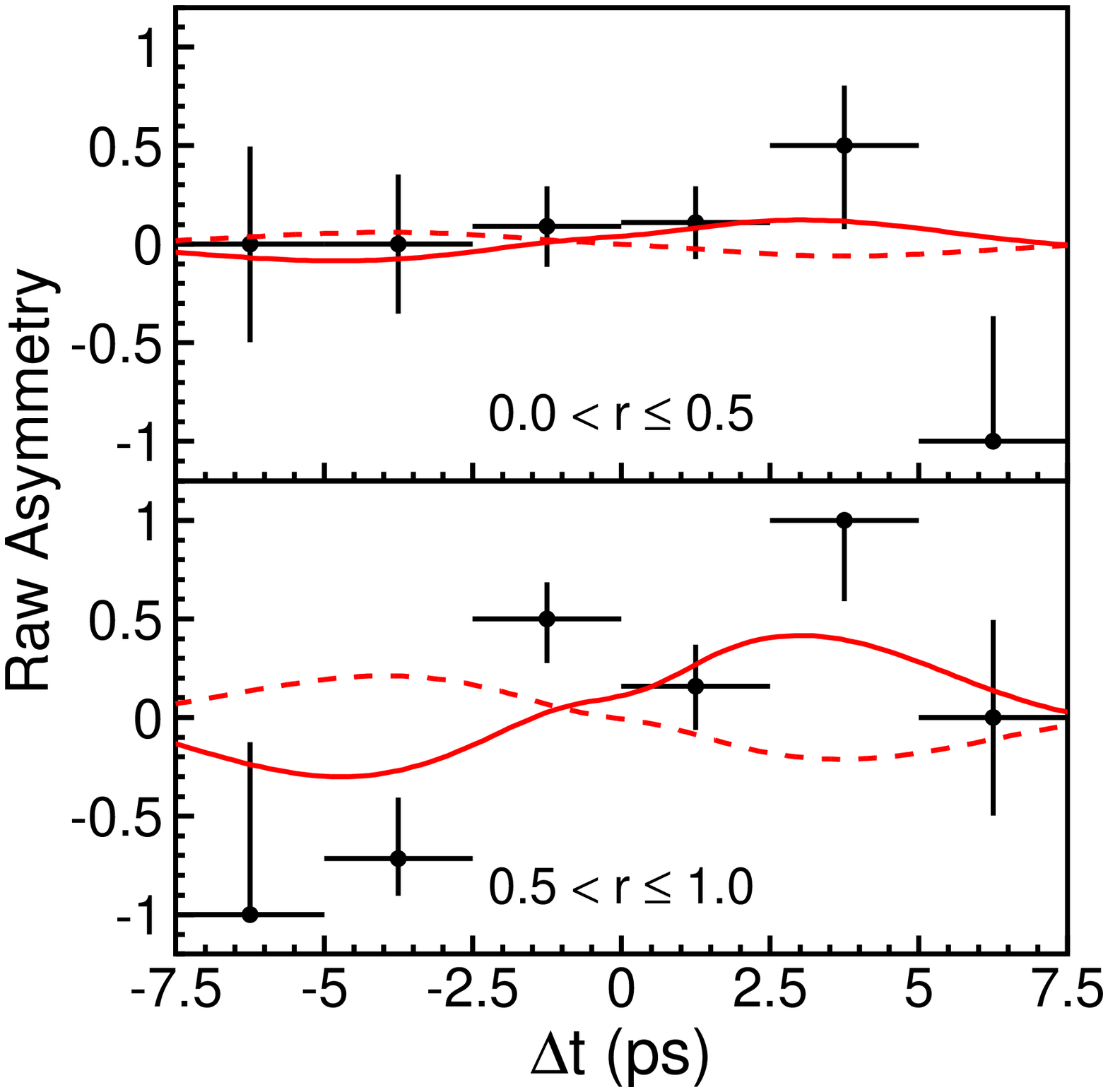}} 
\resizebox{!}{0.49\textwidth}{\includegraphics{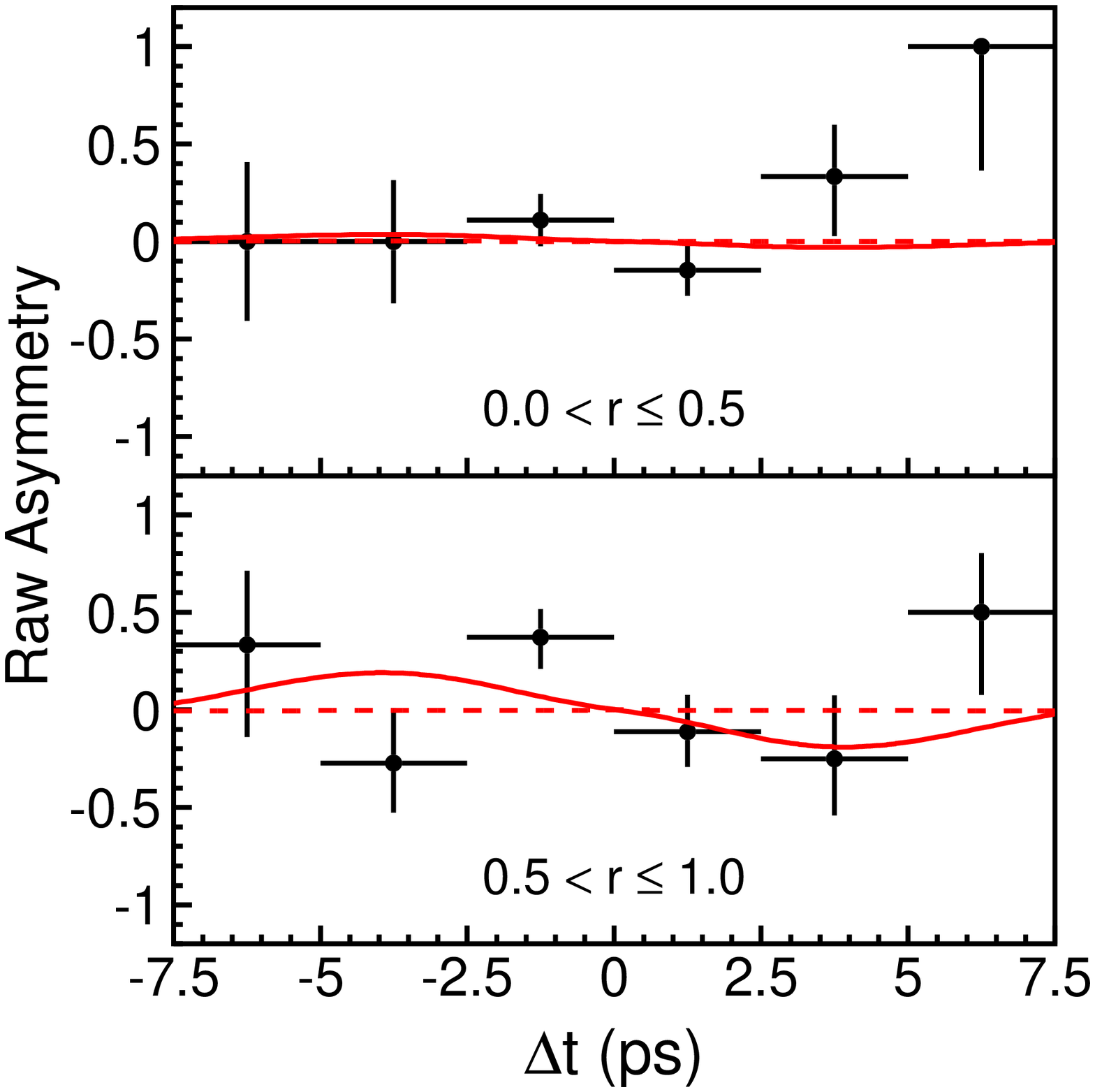}} 
\caption{
The asymmetry, $A$, in each $\Dt$ bin with $0 < r \le 0.5$ (top)
and with $0.5 < r \le 1.0$ (bottom) for 
(a) $\bz\to\ks\ks\ks$ and 
(b) $\bz\to\ks\piz\gamma$.
The solid curves show the result of the 
unbinned maximum-likelihood fit.
The dashed curves show the SM expectation with 
$(\cals, \cala) = (-\sin2\phi_1 = -0.73, 0)$ 
for $\bz\to\ks\ks\ks$ and
with $(\cals, \cala) = (0, 0)$ 
for $\bz\to\ks\piz\gamma$.}
\rput[l](-6.5, 9.7)  {(a)~$\ks\ks\ks$}
\rput[l]( 1.8, 9.7)  {(b)~$\ks\piz\gamma$}
\label{fig:asym}
\end{figure}

Table \ref{tab:result} summarizes
the fit results of $\cals$ and $\cala$.
We define the raw asymmetry in each $\Dt$ bin by
$(N_{q=+1}-N_{q=-1})/(N_{q=+1}+N_{q=-1})$,
where $N_{q=+1(-1)}$ is the number of 
observed candidates with $q=+1(-1)$.
Figures~\ref{fig:asym}(a) and (b)
show the raw asymmetries for the $\bz\to\ks\ks\ks$ and
$\ks\piz\gamma$ decays, respectively,
in two regions of the flavor-tagging
parameter $r$. While the numbers of events in the two regions are similar,
the effective tagging efficiency is much larger 
and the background dilution is smaller in the region $0.5 < r \le 1.0$.
Note that these projections onto the $\Delta t$ axis do not take into
account event-by-event information (such as the signal fraction, the
wrong tag fraction and the vertex resolution), which is used in the
unbinned maximum-likelihood fit.

Table~\ref{tab:syserr}
lists the systematic errors on $\cals$ and $\cala$.
The total systematic errors are obtained
by adding each contribution in quadrature,
and are much smaller than the statistical errors for all modes.

To determine the systematic error that arises from
uncertainties in the vertex reconstruction,
the track and vertex selection criteria
are varied to search for possible systematic biases.
Small biases in the $\Dz$ measurement 
are observed in $e^+e^-\to\mu^+\mu^-$ and other control
samples. Systematic errors 
are estimated by applying special correction functions
to account for the observed biases, repeating
the fit, and comparing the obtained values with the nominal results.
The systematic error due to the IP constraint 
in the vertex reconstruction is estimated by
varying ($\pm10 \mu$m) the smearing used to account for the
$B$ flight length.
Systematic errors due to imperfect SVD alignment
are determined
from MC samples that have artificial mis-alignment effects
to reproduce impact-parameter resolutions observed in data.

Systematic errors due to uncertainties in the wrong tag
fractions are studied by varying
the wrong tag fraction individually for each $r$ region.
Systematic errors due to uncertainties in the resolution function
are also estimated by varying each resolution parameter obtained from
data (MC) by $\pm 1\sigma$ ($\pm 2\sigma$), repeating the fit
and adding each variation in quadrature.
Each physics parameter such as $\taubz$ and $\dmd$
is also varied by its error.
A possible fit bias is examined by fitting a large number of MC events.

Systematic errors from uncertainties in the background fractions
and in the background $\Dt$ shape
are estimated by varying each background parameter obtained
from data (MC) by $\pm 1\sigma$ ($\pm 2\sigma$).
Uncertainties in the background $B$ decay model
are also considered for
the $\bz\to\ks\piz\gamma$ mode;
we compare different theoretical models for radiative $B$ decays
and take the largest variation as the systematic error.

Additional sources of systematic errors are 
considered for $B$ decay backgrounds
that are neglected in the PDF.
We consider uncertainties both in their fractions
and $CP$ asymmetries; for modes that have
non-vanishing $CP$ asymmetries, we conservatively
vary the $CP$-violation parameters within the
physical region and take the largest variation
as the systematic error.
The effect of backgrounds from $\chi_{c0}\ks$
in the $\bz\to\ks\ks\ks$ sample is considered.

Finally, we investigate the effects of interference between
CKM-favored and CKM-suppressed $B\to D$ transitions in
the $\ftag$ final state~\cite{Long:2003wq}.
A small correction to the PDF for the signal distribution
arises from the interference.
We estimate the size of the correction using the $\bzdslnu$ 
sample. We then generate MC pseudo-experiments
and make an ensemble test to obtain systematic biases
in $\cals$ and $\cala$.

Various crosschecks of the measurement are performed.
We reconstruct charged $B$ meson decays
that are the counterparts of the $\bz\to\fCP$ decays
and apply the same fit procedure.
All results for the $\cals$ term are consistent with no 
$CP$ asymmetry, as expected. 
Lifetime measurements are also performed for 
the $\fCP$ modes and the corresponding charged $B$ decay modes.
The fits yield
$\taubz$ and $\taubp$ values consistent with the world-average values.
MC pseudo-experiments are generated for each decay mode to
perform ensemble tests.
We find that the statistical errors obtained
in our measurements are all consistent
with the expectations from the ensemble tests.

A fit to the $\bz\to\ks\piz\gamma$ subsample 
with $0.6 < M_{\ks\piz} < 0.8$ GeV/$c^2$ or
$1.0 < M_{\ks\piz} < 1.8$ GeV/$c^2$, which
excludes the $\kstarz$ mass region, 
yields
$\cals = -0.39^{+0.63}_{-0.52}$(stat) and
$\cala = +0.10\pm 0.51$(stat).
The result on the $\cals$ term
is consistent with the previous result obtained
for the decay $\bz\to\kstarz\gamma$~($\kstarz\to\ks\piz$),
$\cals = \SkstarzgmResultSS$,
where 
$0.8 < M_{\ks\piz} < 1.0$ GeV/$c^2$ was required.

\begin{figure}
\resizebox{0.65\textwidth}{!}{\includegraphics{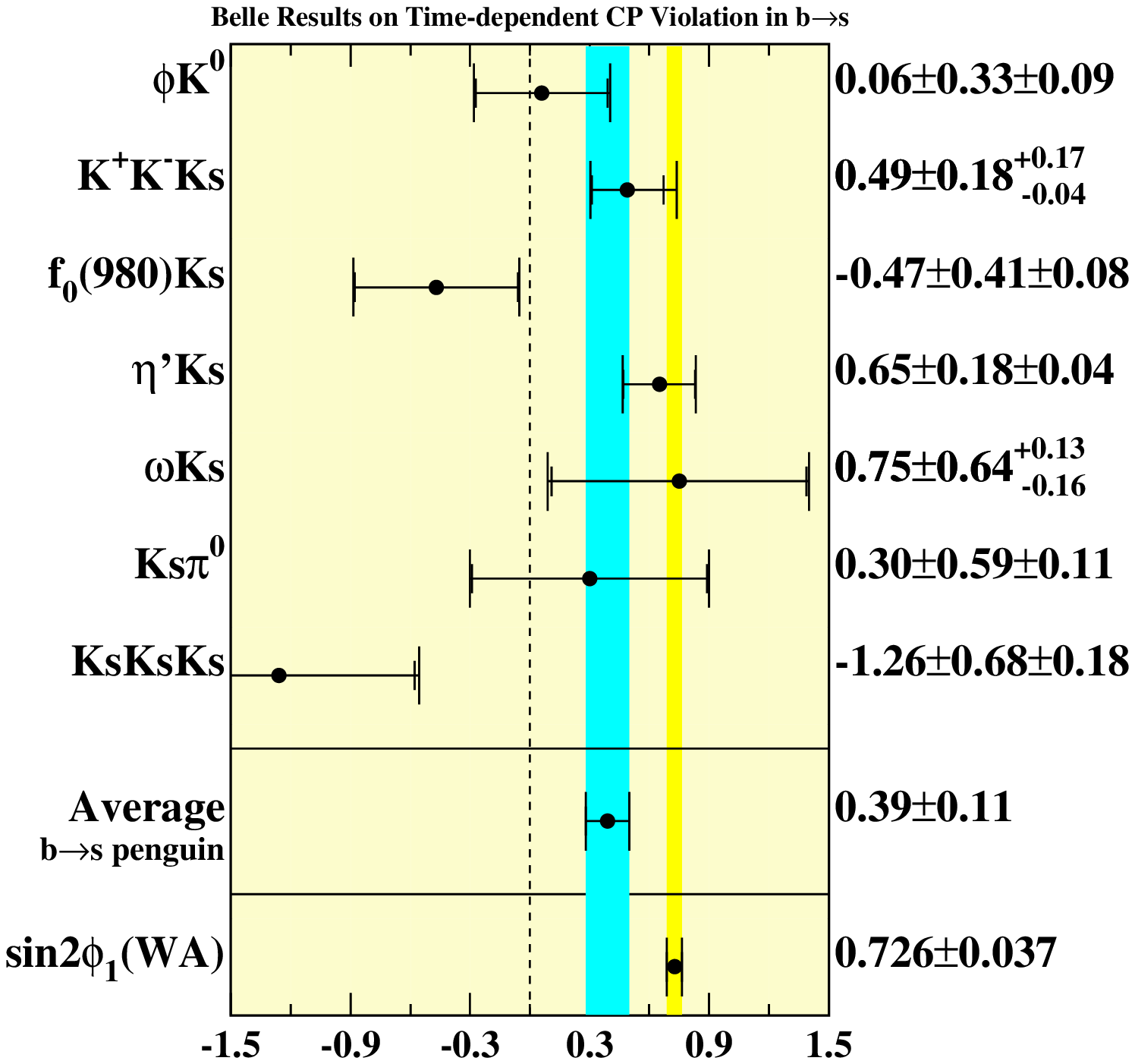}} 
\caption{Summary of $\sinbb$ measurements performed with
$\bz$ decay governed by the $b\to s \overline{q}q$ transition. 
The world-average $\sinbb$ value obtained from $\bz\to\jpsi\kz$
and other related decay modes governed by the $b\to c\overline{c}s$
transition~\cite{bib:HFAG} is also shown as the SM reference.}
\label{fig:avg}
\end{figure}

As discussed in Section~\ref{sec:introduction},
to a good approximation,
the SM predicts $\cals = -\xi_f\sin 2\phi_1$
for the $\bz\to\ks\ks\kz$ decay.
Figure~\ref{fig:avg} summarizes 
the $\sinbb$ determination based on Belle's $\cals$ measurements
using modes dominated by the $b\to s$ transition~\cite{Abe:2004xp}.
For each mode, the first error shown in the figure
is statistical and the second error
is systematic. 
We obtain $\sinbb = \SbsqqNewResult$ as a weighted average,
where the error includes both statistical and systematic errors.
The result differs from the SM expectation by
2.7 standard deviations.

\section{Summary}
We have performed a new measurement of 
$CP$-violation parameters for 
$\bz \to \ks\ks\ks$ decay
based on a sample of $275\times 10^6$ $B\bbar$ pairs.
The decay is dominated by the $b\to s$ flavor-changing
neutral current and
the $\ks\ks\ks$ final state is a $CP$ eigenstate.
Thus it is sensitive to a new $CP$-violating phase
beyond the SM. The result differs from the SM expectation
by 2.8 standard deviations.
The combined result with the decays
$\bz\to\phi\kz$, $\kp\km\ks$, $\fzero\ks$, $\eta'\ks$,
$\omega\ks$ and $\ks\piz$, which are also dominated by
the $b\to s$ transitions,
differs from the SM expectation by 2.7 standard deviations.

We have also measured the time-dependent $CP$ asymmetry
in the decay $\bz\to\ks\piz\gamma$, which
is also sensitive to physics beyond the SM.
The invariant mass of the $\ks\piz$ system is required
to be between 0.6 and 1.8 GeV/$c^2$, which
is an extension of the previous analysis performed
with the decay $\bz\to\kstarz\gamma~(\kstarz\to\ks\piz)$.
The statistical error is much reduced from
the previous analysis by including more events, and
the result is consistent with the previous analysis.

In both cases,
measurements with a much larger data sample are required to 
conclusively establish the existence of a new $CP$-violating phase
beyond the SM.

\section*{Acknowledgments}
We thank the KEKB group for the excellent operation of the
accelerator, the KEK Cryogenics group for the efficient
operation of the solenoid, and the KEK computer group and
the National Institute of Informatics for valuable computing
and Super-SINET network support. We acknowledge support from
the Ministry of Education, Culture, Sports, Science, and
Technology of Japan and the Japan Society for the Promotion
of Science; the Australian Research Council and the
Australian Department of Education, Science and Training;
the National Science Foundation of China under contract
No.~10175071; the Department of Science and Technology of
India; the BK21 program of the Ministry of Education of
Korea and the CHEP SRC program of the Korea Science and
Engineering Foundation; the Polish State Committee for
Scientific Research under contract No.~2P03B 01324; the
Ministry of Science and Technology of the Russian
Federation; the Ministry of Education, Science and Sport of
the Republic of Slovenia; the National Science Council and
the Ministry of Education of Taiwan; and the U.S.\
Department of Energy.




\clearpage
\newpage

\begin{table*}
  \caption{The event fractions $\epsilon_l$,
    wrong-tag fractions $w_l$, wrong-tag fraction differences $\dwl$,
    and average effective tagging efficiencies
    $\eeff^l = \epsilon_l(1-2w_l)^2$ for each $r$ interval for the DS-II.
    The errors for $w_l$ and $\dwl$
    include both statistical and systematic uncertainties.
    The event fractions are obtained from $\jpsi\ks$ data.}
  \begin{ruledtabular}
    \begin{tabular}{ccclll}
      $l$ & $r$ interval & $\epsilon_l$ &\multicolumn{1}{c}{$w_l$} 
          & \multicolumn{1}{c}{$\dwl$}  &\multicolumn{1}{c}{$\eeff^l$} \\
      \hline
 1 & 0.000 -- 0.250 & $0.397\pm 0.015$ & $0.464\pm0.007$ &$+0.010\pm0.007$ &$0.002\pm0.001$ \\
 2 & 0.250 -- 0.500 & $0.146\pm 0.009$ & $0.321\pm0.008$ &$-0.022\pm0.010$ &$0.019\pm0.002$ \\
 3 & 0.500 -- 0.625 & $0.108\pm 0.008$ & $0.224\pm0.011$ &$+0.031\pm0.011$ &$0.033\pm0.004$ \\
 4 & 0.625 -- 0.750 & $0.107\pm 0.008$ & $0.157\pm0.010$ &$+0.002\pm0.011$ &$0.051\pm0.005$ \\
 5 & 0.750 -- 0.875 & $0.098\pm 0.007$ & $0.109\pm0.009$ &$-0.028\pm0.011$ &$0.060\pm0.005$ \\
 6 & 0.875 -- 1.000 & $0.144\pm 0.009$ & $0.016\pm0.005$ &$+0.007\pm0.007$ &$0.135\pm0.009$ \\
    \end{tabular}
  \end{ruledtabular}
\label{tab:wtag} 
\end{table*}
\begin{table}
\caption{
The estimated signal purity and
the signal yield $\nsig$ in the signal region for
$\bz\to\ks\ks\ks$ and $\ks\piz\gamma$ decays.
For the $\bz\to\ks\ks\ks$ decay,
results both before and after vertex reconstruction (
the latter in parentheses)
are shown as events that do not have vertex information
are also used to extract the direct $CP$ violation parameter
$\cala$. 
For the $\bz\to\ks\piz\gamma$ decay, the result
after vertex reconstruction is shown.
}
\label{tab:num}
\begin{ruledtabular}
\begin{tabular}{lll}
\multicolumn{1}{c}{Mode}
                  & \multicolumn{1}{c}{purity} 
                  &\multicolumn{1}{c}{$\nsig$}\\
\hline
$\ks\ks\ks$       & $\PBksksks$          ($\PAksksks$) 
                  & $\NBsigksksks$       ($\NAsigksksks$)\\
$\kspm\kspm\kspm$ & $\PBkspmkspmkspm$    ($\PAkspmkspmkspm$)
                  & $\NBsigkspmkspmkspm$ ($\NAsigkspmkspmkspm$)\\
$\kspm\kspm\kszz$ & $\PBkspmkspmkszz$    ($\PAkspmkspmkszz$)
                  & $\NBsigkspmkspmkszz$ ($\NAsigkspmkspmkszz$)\\
\hline
$\ks\piz\gamma$   & $\Pkspizgm$          
                  & $\Nsigkspizgm$ \\
\end{tabular}
\end{ruledtabular}
\end{table}
\begin{table}
\caption{Results of the fits to the $\Dt$ distributions.
The first error is statistical and the second
error is systematic.}
\label{tab:result}
\begin{ruledtabular}
\begin{tabular}{lrll}
\multicolumn{1}{c}{Mode} &  
SM expectation for $\cals$ &
\multicolumn{1}{c}{$\cals$} & 
\multicolumn{1}{c}{$\cala$} \\
\hline
$\ks\ks\ks$ & $-\sinbb$   & $\SksksksResult$  & $\AksksksResult$ \\
$\ks\piz\gamma$ 
            & $-2(m_s/m_b)\sinbb$ 
                          & $\SkspizgmResult$ & $\AkspizgmResult$ \\
\end{tabular}
\end{ruledtabular}
\end{table}
\begin{table*}
  \caption{Summary of the systematic errors on $\cals$ and $\cala$.}
  \begin{ruledtabular}
    \begin{tabular}{lrrrr}
           & $\cals_{\ks\ks\ks}$ 
                    & $\cals_{\ks\piz\gamma}$ 
                           & $\cala_{\ks\ks\ks}$ 
                                   & $\cala_{\ks\piz\gamma}$ \\
\hline
Vertex 
reconstruction & 0.02 & 0.05 & 0.05 & 0.06\\
Flavor tagging & 0.04 & 0.01 & 0.01 & 0.01\\
Resolution 
function       & 0.12 & 0.05 & 0.04 & 0.04\\
Physics 
parameter      & 0.01 & 0.01 & 0.01 & 0.01\\
Possible 
fit bias       & 0.03 & 0.02 & 0.02 & 0.01\\
Background 
fraction       & 0.10 & 0.06 & 0.03 & 0.04\\
Background 
$\Dt$ shape    & 0.08 & 0.05 & 0.01 & 0.04\\
Tag-side 
interference   & 0.02 & 0.01 & 0.02 & 0.06\\
\hline      
Total          &$\SksksksSyst$ 
                      &$\SkspizgmSyst$
                              &$\AksksksSyst$
                                      &$\AkspizgmSyst$ \\
    \end{tabular}
  \end{ruledtabular}
\label{tab:syserr} 
\end{table*}

\clearpage
\newpage
\clearpage
\newpage
%

%

\begin{thebibliography}{999}
\bibitem{bib:CC}
Throughout this paper,
the inclusion of the charge conjugate decay mode is implied
unless otherwise stated.
\end{thebibliography}
\end{document}